\documentclass[prb,twocolumn]{revtex4-1}
\usepackage{graphicx}
\usepackage{xcolor}
\usepackage{hyperref}
\usepackage{fullpage}

\begin{document}

\title{Anomalous specific heat peak of BaVS$_3$ at 69 K explained by enhanced electron scattering upon ion relocation}
\author{Ferenc M\'{a}rkus}
\email{Corresponding author: markus@phy.bme.hu}
\affiliation{Department of Physics, Budapest University of Technology and Economics, P.O. Box 91, H-1521 Budapest, Hungary}

\author{Bence G. M\'{a}rkus}
\affiliation{Department of Physics, Budapest University of Technology and Economics and MTA-BME
Lend\"ulet Spintronics Research Group (PROSPIN), P.O. Box 91, H-1521 Budapest, Hungary}

\begin{abstract}
The present study deals with the anomalous heat capacity peak and thermal conductivity of BaVS$_3$ near the metal-insulator transition present at $69$ K. The transition is related to a structural transition from an orthorhombic to monoclinic phase. Heat capacity measurements at this temperature exhibit a significant and relatively broad peak, which is also sample dependent. The present study calculates the entropy increase during the structural transition and we show that the additional entropy is caused by enhanced electron scattering as a result of the structural reorientation of the nuclei. Within the model it is possible to explain quantitatively the observed peak alike structure in the heat capacity and in heat conductivity.
\end{abstract}

\maketitle

\section{Introduction}

Materials, which bear the ABX$_3$ structure are still in the center of interest, due to their novel properties, which bear with potential applications, for example MOSFET device fabrication\cite{Neven2004}. One member of this family is barium vanadium sulfide (BaVS$_3$), which has unique electronic properties, such as metal-insulator transition \cite{Imai1996,Fagot2005_1,Fagot2005_2,Ivek2008,Demko2010}, "bad metal" behavior \cite{Kezsmarki2005}, magnetic field induced structural transition \cite{Fazekas2007}, charge density waves (CDW) \cite{Ivek2008}, and is still in the center of interest. It is experimentally \cite{Imai1996,Fagot2005_1,Fagot2005_2,Ivek2008,Demko2010} and theoretically \cite{Fagot2003, Lechermann2007} verified that the barium vanadium sulfide (BaVS$_3$) has an orthorhombic to monoclinic structural transition during the metal-insulator (MI) phase transition at the temperature approximately $T_{\text{MI}} = 69$ K. The given structural transition relates to an extensive regime of one-dimensional lattice fluctuations. Detailed measurements \cite{Imai1996,Demko2010} are elaborated to determine the changes in the measurable physical quantities, like thermal conductivity and specific heat. Since the specific heat is a sensitive indicator of phase transitions it is worth to focus on its behavior. A significant, but relatively broad peak appears at $T_{\text{MI}}$ in the specific heat measurements, as it can be recognized from Fig. 2. and 4. of Ref. \onlinecite{Imai1996} and Ref. \onlinecite{Demko2010}, respectively. This peak is absent in general in MI transitions. It is widely concerned that the reasons of the peak might be explained by the Mott transition via the Brinkman--Rice effect and the contribution of the volumetric change, considered as a Kondo insulator or spin pairing effect \cite{Nakamura1994,Graf1995,Nishihara1981}. The charge density waves near the Peierls transition \cite{Kwok1989,Smontara1993}, charge density and spin wave \cite{Maki1992} are also good candidates in the explanation, however, the caused effect of the previously mentioned phenomena is too small, even when their effects are summed up. Until now, no adequate explanation exists for this behavior. The breadth of the peak has sample quality (e.g. S-component ratio \cite{Shiga1998}) and size dependence.  
 
\section{The entropy increase in the metal-insulator transition}

According to self-consistent electronic structure calculations\cite{Solovyev1994} the specific heat of the metallic regime of BaVS$_3$ agrees well with that of the insulating BaTiS$_3$. The difference of the specific heat among the samples can be calculated, as shown in Fig. 3. of Ref. \onlinecite{Imai1996}. Furthermore, from the obtained curve the extra entropy can be also extracted, presented in Fig. 4 of Ref. \onlinecite{Imai1996}. From the resulted plot the authors claim that, the steepness of the curve is the highest at around the MI transition point, $T_{\text{MI}}=69$ K, and a $4.7$ J$/$K mol extra entropy difference is also extracted from the curve in the range of $40-100$ K.

Later on the authors conclude, that this entropy increase is rather close to the value $R \ln (2S + 1) = 5.76$ J/K mol with $S = 1/2$, which would mean that the degree of freedom for $S = 1/2$ spins has a contribution just above the metal-insulator transition. However, the difference between the experiment and the expected increment from a spin half excitation is more than $22\%$, which is rather large compared to the error of the measurements. Furthermore, the excess entropy in case of the other sample\cite{Demko2010} is a factor of $2$ bigger, than the one observed before and it is clear that this can not be explained by the upper mentioned argument.

\begin{figure}[h!]
    \centering
    \includegraphics[width=0.9\linewidth]{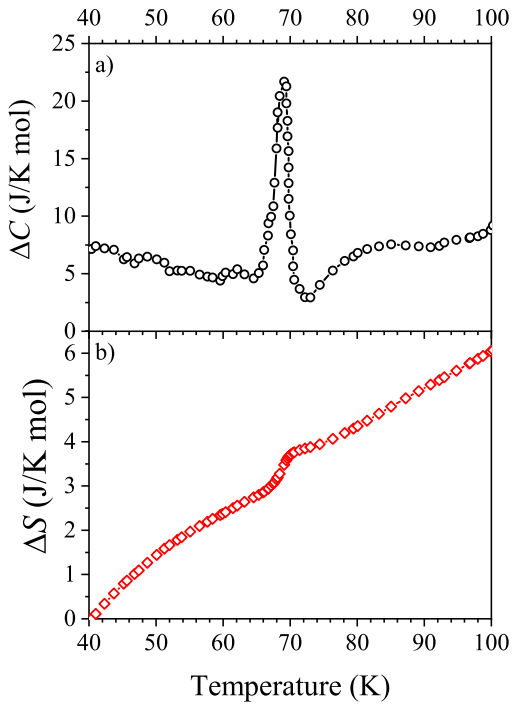}  
    \caption{a) The difference of specific heat near the metal-insulator transition between the two BaVS$_3$ samples presented in Ref. \onlinecite{Demko2010} and in Ref. \onlinecite{Imai1996}. b) Extra entropy in the temperature range of $40-100$ K extracted from the specific heat difference.}
    \label{fig1:delta_c_s}
\end{figure}

Comparing the plots of specific heat in Fig. 4. of Demk\'o \emph{et al.}\cite{Demko2010} it can be seen clearly that the peak observed in the heat capacity is more dominant, pronounced but also sharper, narrower than the one present in the previous article\cite{Imai1996}. Extracting the empirical data from plot, and adopting the idea presented in Ref. \onlinecite{Imai1996}, the difference of specific heat among the two measurements on BaVS$_3$ is plotted in Fig. \ref{fig1:delta_c_s} a) in the relevant temperature range of $40-100$ K. In Fig. \ref{fig1:delta_c_s}b) the explicit entropy excess between the two samples is calculated.
The additional extra entropy is $6.0$ J/K mol between the two BaVS$_3$ samples in the range of $40$ to $100$ K. Compared to the reference BaTiS$_3$ one sample has $4.7$ J/K mol excess entropy at $100$ K, \cite{Imai1996} while the other presents an excess of $10.7$ J/K mol.\cite{Demko2010} This significant difference yields, that the reason of extra entropy must be an additional or a different process.

\section{Contribution of internal electron scattering to the specific heat peak}

The finite width of the peak may suggest oscillations and scattering of the electrons around the ions during the structural transition \cite{Imai1996}, which might be related to the dynamical behavior of the process. However, the damped oscillation of the ion cores, e.g. the acoustic phonon modes can fairly contribute to the internal energy. On the other hand, when electron scattering is taken into account, the core rearrangement can induce an enhancement in the electron scattering rate. We suggest that the effect is arising because of the structural transition and the dramatic change of the band-structure. The mechanism, presented here, takes a model, where a free electron gas is present -- as approaching from the metallic phase -- with the added oscillations arising from core relocation upon phase transition. During the process the internal energy and the specific heat is calculated. The quantitative model reflects well that these oscillations may produce such temperature dependence of specific heat as it is measured.

\begin{figure}[h!]
    \centering
    \includegraphics[width=\linewidth]{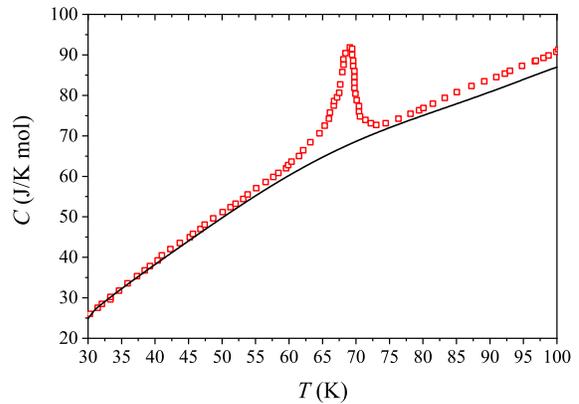}  
    \caption{Red squares are the experimental data presented in Ref. \onlinecite{Demko2010}, black curve is the fitted specific heat with the additional peak at $T_{\text{MI}}$ subtracted.}
    \label{fig2:specific_heat_without_scattering}
\end{figure}

The specific heat curve, measured in Ref. \onlinecite{Demko2010}, is presented without the scattering effect, e.g. without the additional peak in Fig. \ref{fig2:specific_heat_without_scattering} in the temperature range of $30-100$ K. This fitted curve involves the specific heat related to the phonons below the transition temperature of $69$ K. Above the transition temperature it consists the specific heat contribution of phonons and conducting electrons. 
The energy contribution of the electron scattering is approximated by a Lorentzian energy distribution
\begin{equation}
    g(\varepsilon) = \frac{\Gamma/2}{(\varepsilon_{\text{r}} - \varepsilon)^2 + (\Gamma/2)^2},
\end{equation}
where $\Gamma$ is the half width of the distribution, which is related to the scattering rate or the inverse momentum lifetime, $\varepsilon_{\text{r}}$ is the temperature dependent resonant energy and $\varepsilon$ is the instant electron energy \cite{Solyom2009}. It is assumed that the resonance has a maximum at the metal-insulator (structural) transition, e.g. where $T_{\text{res}} = T_{\text{MI}}$. From the width of the specific heat peak it can be physically assumed that the electrons can contribute to the effect from a wider temperature range around the resonant temperature. Thus a relevant Gaussian distribution function for the resonant energy
\begin{equation}
    \varepsilon_{\text{r}} = \varepsilon_0 \, \textrm{exp} \left(-\varepsilon_1 (T-T_{\text{res}})^2 \right)
\end{equation}
can adequately express the physical situation. The parameter $\varepsilon_1$ controls the width of the peak, and expresses that the scattering effect is going below smaller and above higher temperature than the transition temperature. Thus the scattering distribution is a function of temperature as well, $g(\varepsilon, T)$. 

\begin{figure}[h!]
    \centering
    \includegraphics[width=0.9\linewidth]{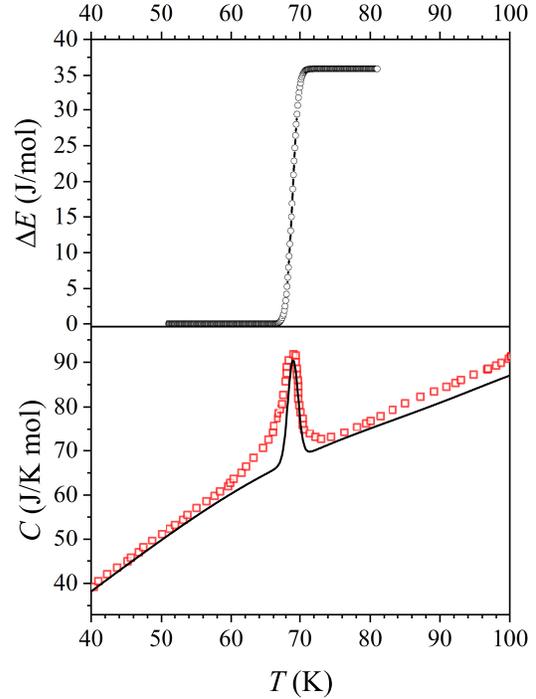}  
    \caption{a) The temperature dependence of energy increase due to internal electron scattering. b) Calculated specific heat of BaVS$_3$. Please note that the baseline, which contains the phononic (and electronic for the metallic regime) contribution is fitted to the experimental data and only the additional peak caused by the enhanced scattering process is calculated.}
    \label{fig3:energy_increase_BaVS3}
\end{figure}

The calculation of the energy increase due to the scattering starts from a reference temperature $T_{\text{ref}}$ to the maximal temperature $T_{\text{max}}$, presently, in our calculations $30$ K and $100$ K. The difference is denoted by $T_{\text{up}} = T_{\text{max}} - T_{\text{ref}}$. The generated energy increase can be calculated by temperature steps $N$, taking into account the continuous change of temperature. Here, we calculate the energy change between the  $T_{i} - T_{i+1}$ as
\begin{equation}
    \Delta\varepsilon_i (T_i) = \frac{N_0}{N} \int\limits_{0}^{\infty} \varepsilon~g(\varepsilon,T_i) f(\varepsilon,T_i)~ \textrm{d} \varepsilon,
\end{equation}
here $f$ is the Fermi-Dirac distribution, furthermore,
\begin{equation}
    T_i = T_{\text{ref}} + \frac{i}{N}T_{\text{up}}
\end{equation}
is the instantaneous temperature between the range $T_{\text{ref}} < T_i <T_{\text{max}}$ using the notion $i =0, 1, 2, \dots, N$. The total energy change is therefore the sum of the $\Delta \varepsilon_i(T_i)$ terms as
\begin{equation}
    \Delta\varepsilon(T) = \sum\limits_{i=1}^{n} \Delta\varepsilon_i(T_i),
\end{equation}
where $T = T_{\text{ref}} + \frac{n}{N}T_{\text{up}}$ with $n \in [n,N]$. The resulted energy increase is plotted in Fig. \ref{fig3:energy_increase_BaVS3}a), from where it can be seen that the internal energy has a stepwise increase at the metal-insulator transition. The specific heat is the temperature derivative of the internal energy
\begin{equation}
    C = \left[ \frac{1}{n} \frac{\partial E}{\partial T} \right]_V,
\end{equation}
where $n$ is the amount of substance. Adding the phononic (and electronic in the case of the metallic regime) contribution to the specific heat from Fig. \ref{fig2:specific_heat_without_scattering} to the derivative of energy increase in Fig. \ref{fig3:energy_increase_BaVS3}a) we obtain a peak at $69$ K, as presented in Fig \ref{fig3:energy_increase_BaVS3}b). The obtained curve agrees well with the experimental results.

The very same calculations can be done for the sample prepared by Imai \emph{et al.}\cite{Imai1996}. The calculated parameters are collected in Table \ref{parameters}. for the two samples.
\begin{table}[h!]
    \centering
    \begin{tabular}{|c|c|c|}
        \hline
        Variables & Sample from Ref. \onlinecite{Demko2010} & Sample from Ref. \onlinecite{Imai1996} \\ \hline
        $\Gamma$ & $6 \times 10^{-23}$ \text{eV} & $60 \times 10^{-23}$ \text{eV}  \\ \hline
        $\varepsilon_0$ & $6~\text{meV}$ & $6~\text{meV}$ \\ \hline
        $\varepsilon_1$ & $0.5~\text{1/K}^2$ & $0.08~\text{1/K}^2$ \\ \hline
    \end{tabular}
\caption{List of calculated parameters for the two different samples.}
\label{parameters}
\end{table}
Here we wish to emphasize, that $\Gamma$ and $\varepsilon_1$ are strongly depend on sample purity, as defects and sample purity clearly cause a higher momentum relaxation rate, which in extreme cases can hinder the observed enhanced electron scattering. In principle these two parameters can also characterize the further samples in terms of purity.

\section{The heat conductivity peak}

An interesting consequence of the scattering effect is a small peak in the heat conductivity at $69$ K as seen in Fig. 4 in Ref. \onlinecite{Demko2010}. In metals the total heat current due to the electrons \cite{Solyom2009} can be written as
\begin{equation}
    J_{\text{q}} = -\frac{1}{3} n_{\text{e}} v l \frac{\partial u}{\partial z},
\end{equation}
where $n_e$ is the electron density, $v$ is the electron speed, $l$ is the mean free path, $u$ is the transported internal energy by one electron and $z$ is the spatial coordinate. Since,
\begin{equation}
    \frac{\partial u}{\partial z} = \underbrace{\frac{\partial u}{\partial T}}_{c} \frac{\partial T}{\partial z},
\end{equation}
where $c$ is the heat capacity per electrons and introducing $C = n_{\text{e}} c$, it is possible to connect heat capacity with heat conductivity by the relation 
\begin{equation}  \label{kappa_heat_cond}
    \kappa = \frac{1}{3} Cvl.
\end{equation}

To calculate the heat conductivity peak caused by the heat capacity peak first an off-resonant value at $66$ K from the graphs in Ref. \onlinecite{Demko2010} has to be taken, where $\kappa_1 = 0.82$ W/K$\cdot$m and $C_1 = 80$ J/K mol = $6.6\times 10^5$J/K m$^3$ can be found. The BaVS$_3$ is known to be a bad metal where the mean free path of electrons is in the order of V$-$V distance, typically $l \approx 0.28$ nm. Using Eq. (\ref{kappa_heat_cond}) the electron speed can be calculated, $v_1 = 1.4 \times 10^4$ m/s. In the second step, values at resonance, $T = 69$ K are taken: $\kappa_2 = 0.84$ W/K$\cdot$m and $C_2 = 92$ J/K mol = $7.6\times 10^5$J/K m$^3$. The obtained electron speed is $v_2 = 1.2 \times 10^4$ m/s, which is equal to $v_1$ within the error of the measurement. This proves that the heat capacity peak is directly related to the peak in heat conductivity and also caused by the enhanced (resonant) electron scattering. 

\section{Conclusion}

Several phenomena may have role in the extra heat capacity of BaVS$_3$ at the temperature $69$ K of metal-insulator transition. Yet, the contribution of these effects is not enough in the explanation of the rather observed peak. The present work shows that an internal electron scattering process due to the structural transition may carry such addition energy that leads to the strong increase of heat capacity around the transition point. The peak in the heat conductivity is also explained.

\section{Acknowledgment}

We wish to thank Prof. Gy\"orgy Mih\'aly and Dr. L\'aszl\'o Demk\'o for highlighting the current topic related to BaVS$_3$ and providing the basic literature. Support by the National Research, Development and Innovation Office of Hungary (NKFIH) Grant Nrs. K119442 and 2017-1.2.1-NKP-2017-00001, and by the BME Nanonotechnology FIKP grant of EMMI (BME FIKP-NAT), are acknowledged.


\begin{thebibliography}{100}
\bibitem{Neven2004} N. Bari\v{s}i\'c, Study of Novel Electronic Conductors: The case of BaVS$_3$, PhD Thesis, Ecole Polytechnique F\'ed\'erale de Lausanne (EPFL), 2004.
\bibitem{Imai1996} H. Imai, H. Wada and M. Shiga, J. Phys. Soc. Japan {\bf 65}, 3460 (1996).
\bibitem{Fagot2005_1} S. Fagot, P. Foury-Leylekian, S. Ravy, J. P. Pouget, M. Anne, G. Popov, M. V. Lobanov and M. Greenblatt, Solid State Sci. {\bf 7}, 718–725 (2005).
\bibitem{Fagot2005_2} S. Fagot, P. Foury, S. Ravy, J. P Pouget, G. Popov, M. V. Lobanov, M. Greenblatt, Physica B {\bf 359–361}, 1306–1308 (2005).
\bibitem{Ivek2008} T. Ivek, T. Vuleti\'c, S. Tomi\'c, A. Akrap, H. Berger and L. Forr\'o, Phys. Rev. B {\bf 78}, 035110 (2008).
\bibitem{Demko2010} L. Demk\'o, I. K\'ezsm\'arki, M. Csontos, S. Bord\'acs and G. Mih\'aly, Eur. Phys. J. B {\bf 74}, 27 (2010).
\bibitem{Kezsmarki2005} I. K\'ezsm\'arki, G. Mih\'aly, R. Ga\'al, N. Bari\v{s}i\'c, H. Berger, L. Forr\'o, C. C. Homes and L. Mih\'aly, Phys. Rev. B {\bf 71}, 193103 (2005). 
\bibitem{Fazekas2007} P. Fazekas, N. Bari\v{s}i\'c, I. K\'ezsm\'arki, L. Demk\'o, H. Berger, L. Forr\'o and G. Mih\'aly, Phys. Rev. B {\bf 75}, 035128 (2007).
\bibitem{Fagot2003} S. Fagot, P. Foury-Leylekian, S. Ravy, J. P. Pouget and H. Berger, Phys. Rev. Lett. {\bf 90}, 196401 (2003).
\bibitem{Lechermann2007} F. Lechermann, S. Biermann and A. Georges, Phys. Rev. B {\bf 76}, 085101 (2007).
\bibitem{Nakamura1994} M. Nakamura, A. Sekiyama, H. Namatame, A. Fujimori, H. Yoshihara, T. Ohtani, A. Misu, and M. Takano, Phys. Rev. B {\bf 49}, 16191 (1994).
\bibitem{Graf1995} T. Graf, D. Mandrus, J. M. Lawrence, J. D. Thompson, P. C. Canfield, S.-W. Cheong, and L. W. Rupp, Jr., Phys. Rev. B {\bf 51}, 2037 (1995).
\bibitem{Nishihara1981} H. Nishihara, M. Takano, J. Phys. Soc. Japan {\bf 50}, 426 (1981).
\bibitem{Kwok1989} R. S. Kwok, S. E. Brown, Phys. Rev. Lett. {\bf 63}, 895 (1989).
\bibitem{Smontara1993} A. Smontara, K. Biljakovic’, S. N. Artemenko, Phys. Rev. B {\bf 48}, 4329 (1993).
\bibitem{Maki1992} K. Maki, Phys. Rev. B {\bf 46}, 7219 (1992).
\bibitem{Shiga1998} M. Shiga, H. Imai, H. Wada, J. Magn. and Magn. Mat. {\bf 177-181}, 1347 (1998).
\bibitem{Solovyev1994} I. V. Solovyev, V. I. Anisimov and E. Z. Kurmaev, Physica Scripta {\bf 50}(1), 90-92 (1994).
\bibitem{Solyom2009} J. S\'olyom, Fundamentals of the Physics of Solids: Volume II - Electronic Properties (Springer, Berlin, 2009).
\bibitem{Kezsmarki2006} I. K\'ezsm\'arki, G. Mih\'aly, R. Ga\'al, N. Bari\v{s}i\'c, A. Akrap, H. Berger, L. Forr\'o, C. C. Homes and L. Mih\'aly, Phys. Rev. Lett. {\bf 96}, 186402 (2006).
\end{thebibliography}
\end{document}